\documentclass[onecolumn,floatfix,superscriptaddress,showpacs,showkeys,nofootinbib,preprint]{revtex4}
\usepackage{epsfig}
\usepackage{amssymb,latexsym,amsmath}

\begin{document}

\title{Determination of the cross sections of (n,2n), (n,$\gamma$) nuclear reactions on germanium isotopes at the energy of neutrons 13.96 MeV}

 \author{S.V. Begun{\footnote {Corresponding author, E-mail: begun@univ.kiev.ua}}}
 \affiliation{Kyiv National Taras Shevchenko University, Kyiv, Ukraine}
 \author{O.G. Druzheruchenko}
 \affiliation{Kyiv National Taras Shevchenko University, Kyiv, Ukraine}
 \author{O.O. Pupirina}
 \affiliation{Institute for Nuclear Research National Academy of
Sciences of Ukraine, Kyiv, Ukraine}
 \author{V.K. Tarakanov}
 \affiliation{Kyiv National Taras Shevchenko University, Kyiv, Ukraine}

\begin{abstract}
The cross sections of $^{70}$Ge(n,2n)$^{69}$Ge,
$^{72}$Ge(n,2n)$^{71}$Ge, $^{76}$Ge(n,$\gamma$)$^{77(g+0.21m)}$Ge,
$^{76}$Ge(n,2n)$^{75}$Ge nuclear reactions were measured at the
energy of neutrons (13.96$\pm$0.06) MeV by activation method with
$\gamma$-ray and X-ray spectra studies.
\end{abstract}



\maketitle

%-----------------------------------------------------------------------
\section{Introduction}
Germanium is the the material for the detectors which can be used
at the measurements of $\gamma$-ray spectra \cite{neutr, Akimov}
or charged particles energy \cite{Akimov} under fast neutrons
background existence. To calculate the contribution of these
neutrons to measured pulse-height spectra one must have a set of
reliable experimental data. Experimental results on nuclear
reaction cross sections in this energy region also are useful for
testing nuclear reaction models. From this point of view there is
no information on experimental measurements of
$^{72}$Ge(n,2n)$^{71}$Ge and $^{76}$Ge(n,$\gamma$) nuclear
reaction cross sections at the considered neutron energy range
\cite{EXFOR}. But these nuclear constants could be measured by
means of activation technique.

%-----------------------------------------------------------------------
\section{Method of the experiment}

The samples of polycrystal germanium, of metallic zirconium and
niobium were examined on purity by preliminary X-ray fluorescent
analysis and by neutron activation method. All the samples are of
natural isotopic composition. Neutrons were generated by the
T(d,n)$^4$He nuclear reaction. Low-voltage accelerator of charged
particles was used. Mixed beam of ions D$^+$ and D$_{2}^{+}$ was
accelerated by 220 kV potential. The diameter of deuteron beam on
the T-Ti target was 25 mm. But the concentration of tritium at the
central part of the target was approximately zero. This fact was
confirmed by comparing of calculated at the assumption of uniform
concentration and measured average energy of neutrons. Another
confirmation of this fact is the experimental and calculated
neutron flux density at the irradiation points. The distance from
the set of samples to T-Ti target was 5.5 mm. The time of the
irradiation of samples was chosen 1.33 hour for considered nuclear
reactions. The average neutron flux density at the irradiation
points was determined by using of niobium samples. The sample of
germanium was placed between two foils of niobium with the same
diameter at the irradiation time. The reference nuclear reaction
for the flux density measurements was $^{93}$Nb(n,2n)$^{92m}$Nb.
The set of samples of niobium and germanium was placed between two
samples of zirconium. The samples of zirconium and niobium were
used for the determination of the average energy of neutrons at
the irradiation points by Zr/Nb method \cite{Zr/Nb, Dis} The set
of samples of niobium, zirconium and germanium at the time of
irradiation were covered by 0.4 mm cadmium foils to reduce the
influence of thermal neutrons. Diameter of germanium, niobium and
zirconium samples was 10 mm or 20 mm. Thicknesses of germanium,
niobium and zirconium samples are 0.7 mm, 0.1 mm, 0.03 mm
accordingly for 10 mm diameter and 3 mm, 0.1 mm, 0.3 mm
accordingly for 20 mm diameter.
\par The spectra of activation products were measured by
$\gamma$-spectrometer with HPGe-detector (sensitive volume
$\sim$110 cm$^3$) and by X-ray spectrometer with planar
Si(Li)-detector (active area $\sim$28 mm$^2$, thickness $\sim$2.5
mm). The distance from the sample to the sensitive area of the
X-ray detector was 15 mm with 4 mm diameter collimator. The
self-absorption correction was calculated by Monte-Carlo
simulation method with taking into account the non-uniform
distribution of activity and the influence of detector edge
effects \cite{Dis, Selfabs}. X-rays spectra of germanium samples
should be measured after 21 days from the irradiation time. This
cool-down time is necessary for $^{72}$Ge(n,2n)$^{71}$Ge nuclear
reaction cross section measurement. After this period of time one
can neglect the activity of all other activation products of
germanium isotopes. The activity of X-rays after this period of
time will be formed only by electron capture decay of $^{71}$Ge.
Measurements of $\gamma$-spectra were carried out at 81 mm
distance to the sensitive volume of HPGe-detector because of high
activity of the irradiated samples. The correction on true
coincidence summing of gamma-quanta and on finite geometry effects
are applied \cite{Dis}. True coincidence summing correction is
calculated by semi-empirical method \cite{Dis}. Besides the
$\gamma$-$\gamma$ true coincidence the influence of X-rays
originating from internal conversion or preceded by electron
capture, positron annihilation photons, absorbed and scattered in
the sample $\gamma$-radiation, bremsstrahlung radiation are
considered. The calculation of the correction on the dead-time of
the spectrometer was performed for the nonstationary conditions of
measurements; to do this the spectrometer indication of dead-time
losses with the a posteriori separated contributions to the total
activity was used. The calculation of primary neutron source
spectra profile was performed with computer code SPECTRON (Khlopin
Radium Institute, St.Petersburg, Russia). The full width at half
maximum of neutron spectrum was derived from these calculations.
The geometry of irradiation and the most considerable parameters
of the neutron generator set and of the neutron generator hall
were taken into account at the calculations of the secondary
neutrons spectra profile by Monte-Carlo method.

%--------------------------------------------------------------------------
\section{Experimental results}

Average neutron energy determined by Zr/Nb method was
(13.96$\pm$0.06) MeV. The calculated full width at half maximum
(FWHM) of the primary neutrons spectrum profile was 0.4 MeV. The
energy resolution in this case was 0.2 MeV=(FWHM)/2. The average
neutron flux density at the irradiation point was
$\sim$3.6$\cdot$10$^8$ cm$^{-2}$$\cdot$s$^{-1}$. Calculated with
evaluated data from JEF-2.2 library correction on the secondary
neutrons for $^{76}$Ge(n,$\gamma$)$^{77(g+0.21m)}$Ge nuclear
reaction was $\sim$0.8. One can conclude based on this value that
the main contribution to the measured activity of $^{77}$Ge is
from primary neutrons. Evaluated by author \cite{Dis} cross
section of reference nuclear reaction $^{93}$Nb(n,2n)$^{92m}$Nb
(456$\pm$6) mb was used. All decay data were taken from
\cite{Decay Data}. Results of the experiment are summarized in the
Table \ref{Results}. The results of other authors were taken from
\cite{EXFOR}. All of our results are in good agreement with recent
experimental results of other authors. There is no information on
experimental measurements of $^{72}$Ge(n,2n)$^{71}$Ge and
$^{76}$Ge(n,$\gamma$) nuclear reaction cross sections at the
considered neutron energy range. This fact gives us the
opportunity to note that these values were measured at first. Our
results for these two nuclear reactions are in good agreement with
evaluated data from most of evaluated libraries. The cross
sections of (n,2n) and (n,$\gamma$) nuclear reactions on other
naturally occurring isotopes of germanium could not be determined
by activation method.

\begin{table}[h]
\begin{center}
\caption{The comparison of our data with the the data of other
authors (in mb)} \label{Results}
\begin{tabular}{|l|l|l|l|}
 \hline
Reaction & T$_{1/2}$ & This work  & Other works $^{1)}$\\
\hline $^{70}$Ge(n,2n)$^{69}$Ge & 39.05 h & 420$\pm$30 & 407$\pm$49 (13.99, Konno, 1993) \\
 & & & 392$\pm$61 (13.9, Hoang, 1992) \\
\hline $^{72}$Ge(n,2n)$^{71}$Ge & 11.43 d & 780$\pm$60 & no results \\
\hline $^{76}$Ge(n,$\gamma$)$^{77(g+0.21m)}$Ge & 11.30 h & 1.01$\pm$0.14 & no results \\
\hline $^{76}$Ge(n,2n)$^{75}$Ge & 82.78 m & 1310$\pm$140 & 1300$\pm$101, (14.10, Molla, 1997) \\
 & & & 1160$\pm$130 (14.00, Steiner, 1969) \\
 \hline
\multicolumn{4}{l}{\footnotesize $^{1)}$ The neutron energy, first
author and year of publication are given in  parenthesis}
\end{tabular}
\end{center}
\end{table}

%\newpage
\begin{acknowledgments}
The work was partially supported by contract 06BF051-05 with Kyiv
National Taras Shevchenko University. The author S.V. Begun has
proposed the method of the experiment, has carried out all the
measurements with $\gamma$-ray spectrometer, has determined the
average energy of neutrons by Zr/Nb method, has performed the
calculation of primary neutron source spectra profile, the
correction on dead-time of the spectrometer, the contribution of
self-absorption effect, the contribution of the effect of true
coincidence summing of $\gamma$-quanta, the contribution of the
effects of finite geometry and the calculation of final values.
The co-authors contribution to this work is next. O.G.
Druzheruchenko has carried out all the measurements with X-ray
spectrometer. O.O. Pupirina has performed the calculation of
secondary neutrons spectra profile. V.K. Tarakanov has carried out
the irradiation of samples by DT-neutrons.

\end{acknowledgments}

\end{document}